\documentclass[twocolumn,english,prb,superscripaddress,showpacs, reprint]{revtex4}
\usepackage[T1]{fontenc}
\usepackage[latin9]{inputenc}
\usepackage{amstext}
\usepackage{graphicx}
\usepackage{amssymb}

\makeatletter

\DeclareRobustCommand{\greektext}{%
  \fontencoding{LGR}\selectfont\def\encodingdefault{LGR}}
\DeclareRobustCommand{\textgreek}[1]{\leavevmode{\greektext #1}}
\DeclareFontEncoding{LGR}{}{}

\newcommand{\lyxmathsym}[1]{\ifmmode\begingroup\def\b@ld{bold}
  \text{\ifx\math@version\b@ld\bfseries\fi#1}\endgroup\else#1\fi}

\providecommand{\tabularnewline}{\\}

\@ifundefined{textcolor}{}
{%
 \definecolor{BLACK}{gray}{0}
 \definecolor{WHITE}{gray}{1}
 \definecolor{RED}{rgb}{1,0,0}
 \definecolor{GREEN}{rgb}{0,1,0}
 \definecolor{BLUE}{rgb}{0,0,1}
 \definecolor{CYAN}{cmyk}{1,0,0,0}
 \definecolor{MAGENTA}{cmyk}{0,1,0,0}
 \definecolor{YELLOW}{cmyk}{0,0,1,0}
 }

\makeatother

\usepackage{babel}

\begin{document}

\title{Structurally driven metamagnetism in MnP and related \emph{Pnma}
compounds}

\author{Z. Gercsi}

\affiliation{Dept. of Physics, Blackett Laboratory, Imperial College London, London
SW7 2AZ UK}

\author{K.G. Sandeman}

\affiliation{Dept. of Physics, Blackett Laboratory, Imperial College London, London
SW7 2AZ UK}
\begin{abstract}
We investigate the structural conditions for metamagnetism in MnP
and related materials using Density Functional Theory. A magnetic
stability plot is constructed taking into account the two shortest
Mn-Mn distances. We find that a particular Mn-Mn separation plays
the dominant role in determining the change from antiferromagnetic
to ferromagnetic order in such systems. We establish a good correlation
between our calculations and structural and magnetic data from the
literature. Based on our approach it should be possible to find new
Mn-containing alloys that possess field-induced metamagnetism and
associated magnetocaloric effects.
\end{abstract}


\maketitle

\section{Introduction\label{sec:Introduction}}

The magnetocaloric effect (MCE) is the change of temperature of a
material in a changing magnetic field, and was first observed in iron
by Emil Warburg \cite{Warburg1881}. It has until now been exploited
in paramagnetic salts as a means of cooling below the temperature
of liquid Helium. However, the high global warming potential of conventional
HFC refrigerants \cite{velders2009} and the large fraction of energy
used for cooling are concerns that currently fuel interest in such
\textquotedbl{}magnetic cooling\textquotedbl{} as a high-efficiency
solid-state method of refrigeration close to room temperature. In
recent years much of the search for viable room temperature magnetic
refrigerants has focused on low cost transition metal-rich compounds
that possess first order magnetoelastic or magnetostructural phase
transitions that can be driven by applied fields of the order of 1~Tesla.
Most of these materials are Mn-containing ferromagnets such as Ni$_{\text{2}}$MnGa~\cite{Pareti2003},
R$_{\text{1-x}}$Mn$_{\text{x}}$MnO$_{\text{3}}$~\cite{Phan2007325},
MnAs~\cite{Wada20021} and MnFe(As,P)~\cite{Tegus2002}. The observation
of MCEs during order-order transitions (in metamagnets) is more rare
and the main examples of large effects near room temperature are in
FeRh~\cite{annaorazov_1992a}, CoMnSi~\cite{SandemanPRB}and Ni$_{\text{2}}$MnSn~\cite{krenke_2005a}.
In some ways metamagnets offer a natural route to the desirably elevated
MCEs associated with a first order, rather than continuous magnetic
transition. However, there are few metamagnets examined for their
MCE that have not already been synthesised for other reasons. In this
paper, we focus on MnP-type (\emph{Pnma} space group) binary and ternary
orthorhombic materials and in particular on what structural factors
support antiferromagnetic order and field-induced metamagnetism. Our
motivitation is to allow prediction of new materials that may have
large MCEs.

The appearance of rich magnetic features with peculiar non-collinear
(NC) magnetism such as fan, helical or cycloidal spin structures observed
in MnP-type binary and the related XMnZ (TiNiSi-type) ternary alloys
has engaged the interest of both theoreticians and experimentalists
for several decades~\cite{nagamiya1967,Kallel1974,Niziol1982}. In
what we term the {}``prototype'' MnP binary alloy, the appearance
of the so-called {}``double-helix'' spiral state has been the subject
of several studies. This is a state in which two pairs of Mn atoms
in each cell take on the same helical wavevector, but with a fixed
phase difference between the pairs. Gribanov and Zavadskii~\cite{Gribanov1983}
ascribed the occurence of the magnetic double helix in MnP-type phases
to the existence of indirect double exchange coupling. Based on an
analysis of the energy stability of helimagnetic spin arrangements,
Fjellvåg et al. discussed the possibility of a metamagnetic transition
by purely carrier localization with no changes in structural parameters.~\cite{Fjellvag198429}.
The complicated crystal structure of MnP is, however, a serious obstacle
to a complete analysis of magnetic exchange integrals based on Goodenough's
double exchange\cite{GoodenoughBook1}. Kallel et al.~\cite{Kallel1974}
concluded that the presence of high ratios of antisymmetric to symmetric
exchange is necessary to observe helical structures. They constructed
a model with exchange integrals from only the nearest and second-nearest
Mn-separations taken into account. Dobrzynski et. al.~\cite{Dobrzynski1989}
argued that it is necessary to consider the relation of exchange interactions
with up to the 7th nearest neighbours in order to account for the
magnetic structure of MnP. In TiNiSi-type ternaries, a similar complexity
emerges. Studies of Co$_{x}$Ni$_{\text{1-x}}$MnGe have similarly
forecast the need to consider up to 8 exchange interactions, especially
when a non-zero moment is present on the Co site~\cite{Niziol1982}.

Despite extensive studies of the formation of non-collinear magnetic
phases in MnP-based materials, no simple, guide-like explanations
have been deduced that may allow the design of new materials with
a field-induced metamagnetic transition near to room temperature.
In this article, we compare our theoretical calculations, which use
density functional theory (DFT), with experimental findings to depict
the most important structural parameters for practical material design.
Our observations of high MCE and giant magneto-elastic coupling in
CoMnSi~\cite{AlexHRPD,SandemanPRB} and the rich variety of materials
in the literature that take on the MnP (or TiNiSi) structure~\cite{landrum1998}
motivate our study of metmagnetism in the MnP class. 

The paper is organised as follows: in sec. \ref{sec:Crystal-structure},
the underlying \emph{Pnma} crystal structure is introduced together
with the details of our computational method. Sec. \ref{sec:MnP-Theory}
describes the results of ab-initio electronic calculations based on
DFT theory. We will show that the ferromagnetic structure of the prototype
MnP binary alloy would undergo a change in its magnetic state as a
result of isodirectional lattice expansion. This change, however can
also be realised by chemical pressure through alloying as is demonstrated
in Sec. \ref{sec:(X)MnP}. Examples from the literature for both 3d
transition metal addition as well as p-block partial replacement in
MnP will be discussed. In Sec. \ref{sec:XMnZ}, we extend this analysis
to other Mn-based ternary (TiNiSi structure) compositions that crystallize
in the \emph{Pnma} space group. Finally, we summarise the practical
factors that can scale the crystal structure into the regime where
a change of magnetic state can be expected.

\section{Crystal structure and Computational Details \label{sec:Crystal-structure} }

\subsection{Crystal Structure }

\begin{figure}
\includegraphics[width=8.5cm]{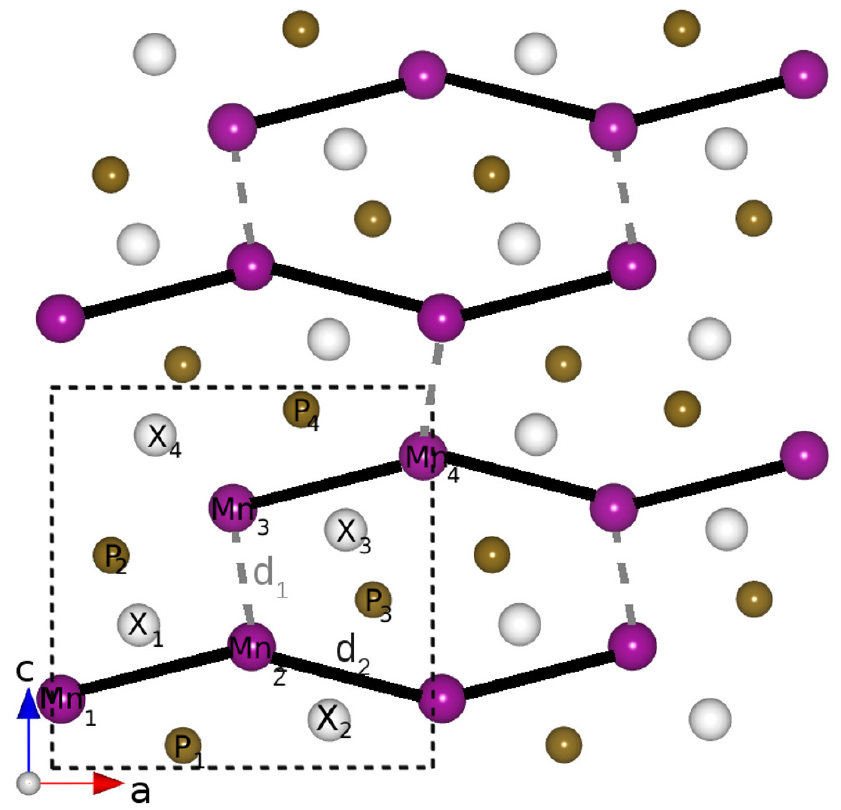}

\caption{\label{fig:MnPstructure} (X)MnP-type (\emph{Pnma}, 62) orthorhombic
structure (the diagram was visualized using VESTA \cite{VESTA}).
The dominant $d_{1}$ and $d_{2}$ Mn-interatomic distances are constructed
with dashed and solid lines, respectively.}

\end{figure}
 The orthorhombic crystal structure of MnP (\emph{Pnma}, 62) is shown
in Fig.\ref{fig:MnPstructure} with unit cell dimensions \emph{a=}5.268~$\lyxmathsym{\AA}$,
\emph{b}=3.172~$\textrm{\AA}$ and c=5.918~$\textrm{\AA}$ at room
temperature. Both the Mn and the P element occupy the 4c ($x,\frac{1}{4},z$)
crystallographic positions with $x_{Mn}=0.0049(2)$, $z_{Mn}=0.1965(2)$
and $x_{P}=0.1878(5)$, $z_{P}=0.5686(5)$~\cite{MnPstructureRudn}.
The orthorhombic structure can be regarded as a distortion from the
higher symmetry hexagonal NiAs (\emph{P}6$_{\text{3}}$/mmc, 194)
type structure with the atomic positions of Ni (4c) shifted from $x_{{\normalcolor Mn}}=0$,
$z_{Ni}=\frac{1}{4}$ to the $x_{Mn}$, $z_{Mn}$ positions and the
As (4c) atoms moved from $x_{Ni}=\frac{1}{4}$, $z_{Ni}=\frac{7}{12}$
to $x_{P}$ and $z_{P}$. Furthermore the two structures can be related
as follows: , $b_{ortho}=a_{hex}$ and $c_{ortho}=\sqrt{3}\times a_{hex}$.
One direct extension of the MnP-type binary structure is the TiNiSi-type
ternary, with the occupation of an additional atomic species at a
general 4c crystallographic position as shown in gray in Fig.~\ref{fig:MnPstructure}.
Therefore in the case of a general XMnP alloy, the number of minimal
basis atoms is 4 for each species in the unit cell and assigned as
Mn$_{1-4}$ and P$_{1-4}$ for a binary with an additional X$_{1-4}$
element in case of a ternary alloy in Fig. \ref{fig:MnPstructure}.
In addition, the two smallest Mn-Mn distances, the focus of our attention
in later sections, are denoted as $d_{1}$ and $d_{2}$.

\subsection{Computational Details}

The electronic structure calculations were carried out using the VASP
code, based on density-functional theory (DFT)\cite{VASP}. Site-based
magnetic moments were calculated using the Voskown analysis within
the general gradient approximation (GGA) scheme. We used the minimal,
8 atom basis supercell for the binary MnP alloy with 4 Mn atoms and
4 P atoms on the 4c crystallographic positions, respectively to calculate
the total energies of the ferro-, antiferro-, and non-magnetic states.
The accurate DoS with high density of k-grid of 21, 19, 21 points
along the a, b and c-axes were used, respectively. The spin\textendash{}orbit
interaction was turned off during the calculations and only collinear
magnetic configurations were considered. The energy convergence criterion
was set to $10^{-7}$~eV during the energy minimization. Where applied,
the expansion and compression of lattice parameters were realized
isotropically.

\section{Results: MnP\label{sec:MnP-Theory}}

As outlined in Section~\ref{sec:Introduction}, MnP exhibits complex
non-collinear ferromagnetic features including low temperature screw
and conical magnetic structures, depending upon the axis of magnetisation.
These transform into collinear ferromagnetism at higher applied fields
and temperatures \cite{MnP_spin1}. %
\begin{table}
\begin{tabular}{|c|c|c|c|c|}
\hline 
Base atoms & $Mn_{1}$ & $Mn_{2}$ & $Mn_{3}$ & $Mn_{4}$\tabularnewline
\hline
Interatomic & \multicolumn{2}{c|}{$d_{2}$} & \multicolumn{2}{c|}{$d_{2}$}\tabularnewline
\cline{2-5} 
distance &  & \multicolumn{2}{c|}{$d_{1}$} & \tabularnewline
\hline 
FM & $\Longrightarrow$ & $\Longrightarrow$ & $\Longrightarrow$ & $\Longrightarrow$\tabularnewline
\hline 
AFM1 & $\Longrightarrow$ & $\Longrightarrow$ & $\Longleftarrow$ & $\Longleftarrow$\tabularnewline
\hline 
AFM2 & $\Longrightarrow$ & $\Longleftarrow$ & $\Longleftarrow$ & $\Longrightarrow$\tabularnewline
\hline 
AFM3 & $\Longrightarrow$ & $\Longleftarrow$ & $\Longrightarrow$ & $\Longleftarrow$\tabularnewline
\hline
\end{tabular}

\caption{\label{tab:MnP_mag-confs} Possible collinear magnetic configurations
of Mn$_{1-4}$in a single unit cell of MnP and their relation to $d_{1}$,$d_{2}$
inter-Mn distances.}

\end{table}
 The collinear ferromagnetic state is stable above $T>47$~K and
it has a Curie temperature of $T_{c}=290$~K \cite{MnP_spin2}. Such
peculiar non-collinear magnetism was explained using spin wave theory
\cite{MnP_spin1,MnP_spin2} and was more recently ascribed to the
influence of magnetocrystalline anisotropy \cite{MnP_anisotropy}.
There have been several theoretical studies of both the magnetism
and crystal symmetry of MnP. The first theoretical work based on augmented
plane wave (APW) theory reported ferromagnetic ground state with moments
of $1.2\mu_{B}$/f.u.\cite{Yanasa_MnP_APW}, in good agreement with
the experimental value of $1.29\mu_{B}$ \cite{MnP_spin1}. Focussing
on the possibility of NiAs-MnP structural distortion, Tremel et. al.
showed the importance of direct and indirect metal-metal bonding due
to a second-order Jahn-Teller type distortion within a tight-binding
(TB) approach~\cite{NiAs-MnPHoffman}. They found that the stabilization
of the lower symmetry MnP phase over the hexagonal one is due to an
increased metal-metal interaction, which results in a lower density
of states at the Fermi level. Spin polarized calculations on the stability
of the orthorhombic phase over the hypothetical NiAs and zincblende-type
ones in MnP was recently also investigated \cite{PRB_MnP_Contineza,PRB_MnP_GGA}
and found good agreement with experimental lattice values.%
\begin{figure}
\includegraphics[width=8.5cm]{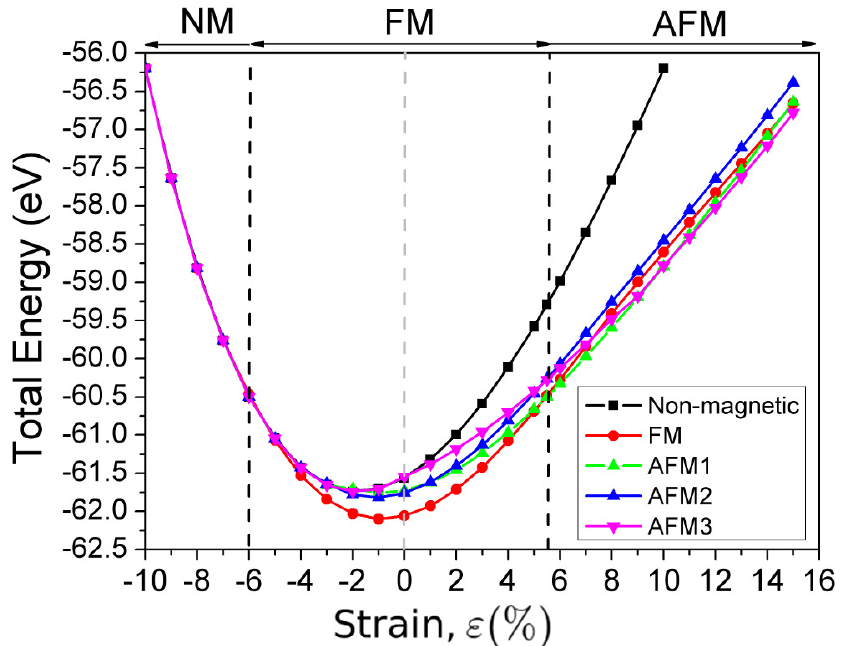}

\caption{\label{fig:Total-energy_MnP} (Color online) Total energy of non-magnetic
(NM), ferromagnetic (FM) and antiferromagnetic (AFM) solutions as
a function of isotropic lattice expansion and compression in MnP.}

\end{figure}
\begin{figure}
\includegraphics[width=8.5cm]{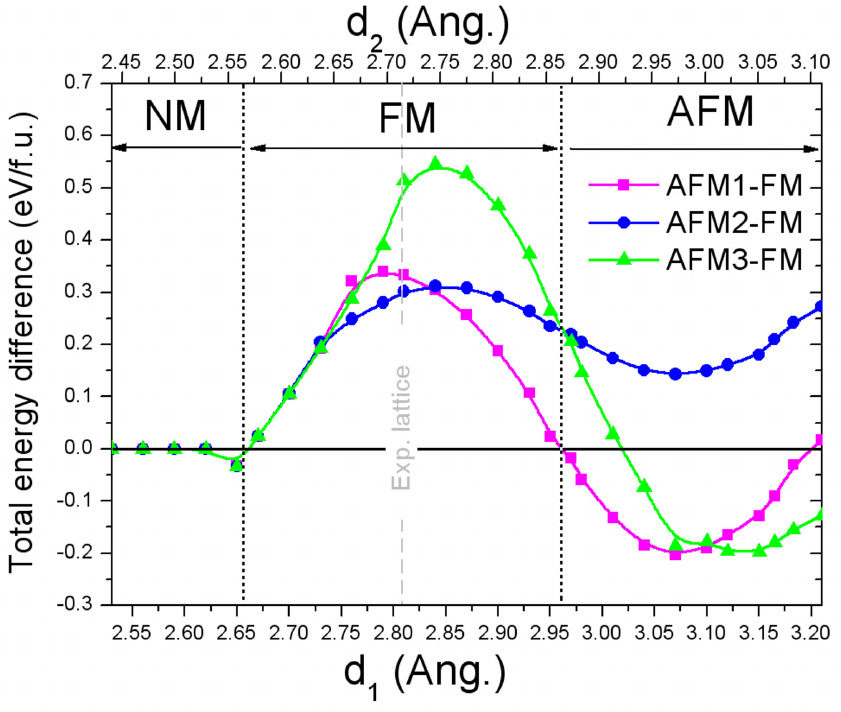}

\caption{\label{fig:Total-energy_MnP-d1d2} (Color online) Difference between
the total energies of antiferromagnetic (AFM) and ferromagnetic (FM)
solutions in MnP as a function of $d_{1}$ and $d_{2}$, the two shortest
Mn-Mn separations. AFM configurations become stable where $\triangle E_{AFM-FM}<0$.
The vertical dashed line represents the experimental (strain-free,
$\varepsilon=0$) Mn-separation values.}

\end{figure}

In this article, we fix the lattice type as \emph{Pnma }as found experimentally
and concentrate on the electronic and magnetic properties of the prototype
MnP binary alloy. We calculate the effect of isotropic lattice expansion
and compression on non-magnetic (NM), ferromagnetic (FM) and antiferromagnetic
(AFM) solutions in order to find the critical lattice parameters where
the crossover from one magnetic state into another can occur. For
this study a single unit cell consisting of 8 atoms (4 Mn and 4 P)
is used, which allows three different collinear antiferromagnetic
configurations (AFM1, AFM2 and AFM3) and a collinear ferromagnetic
(FM) one to be constructed, as show in Table \ref{tab:MnP_mag-confs}. 

Fig. \ref{fig:Total-energy_MnP} shows the calculated total energies
of the converged solutions as a function of isotropic lattice expansion
and shrinkage. As is reflected from the figure, the collinear ferromagnetic
state is predicted to be energetically most favorable in accordance
with experimental observations (above 47~K) in the range~$-0.06\lesssim\varepsilon\lesssim0.05$.
Under severe compression ($\varepsilon\lesssim-0.06$), where the
closest Mn-Mn atomic distance is $d_{2}\lesssim2.55$~Å, no spontaneous
magnetisation is stable in the broad overlapping 3\emph{d} bands and
the calculations find convergence instead in non-magnetic solutions.
On the other hand, with lattice expansion, antiferromagnetic groundstates
can become more stable than the ferromagnetic one, as indicated by
the crossover of the total energy line of AFM1 around $\varepsilon\thickapprox0.05$
in Fig. \ref{fig:Total-energy_MnP}. Further lattice expansion leads
to the AFM3 configuration becoming the most stable one above $\varepsilon\gtrsim0.11$
$(d_{2}\lesssim3.00$~Å). Notably, AFM2 is never favored.

\begin{figure}
\includegraphics[width=8.5cm]{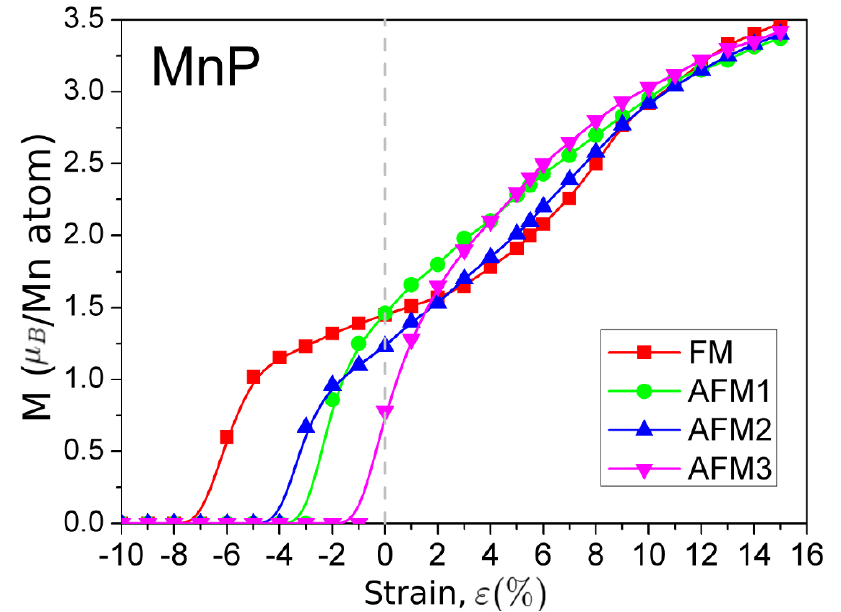}

\caption{\label{fig:MnP_Magn}(Color online) Mn magnetic moment vs. strain
for different magnetic states in MnP (see Table~\ref{tab:MnP_mag-confs}).}

\end{figure}

The relative stability of the different magnetic ground states becomes
more evident if the differences between the energies, \emph{$E_{Tot}$}
of AFM and FM states are compared $(\triangle E_{Tot}=E_{AFM}-E_{FM})$.
By this comparison, a non-FM state becomes most favourable when it
has the most negative value of $\triangle E_{Tot}$. In Fig.~\ref{fig:Total-energy_MnP-d1d2},
the regions of stability of the different magnetic states are shown
as a function of Mn-Mn distance. The FM state is most stable for intermediate
deformations whilst the non-magnetic mode and AFM modes become favored
at high compression and at high strain respectively. Considerable
magnetic moment was found only on the Mn site ($M{}_{Mn}$) as a result
of exchange splitting of the d-orbitals with relatively insignificant
magnetic moments, $M_{P}<0.1\mu_{B}$, on the sites of P atoms. Figure
\ref{fig:MnP_Magn} shows the magnetic moment of Mn atoms in the different
magnetic states as a function of lattice size. In the compressed lattice
region (left part of Figs.~\ref{fig:Total-energy_MnP-d1d2} and \ref{fig:MnP_Magn})
the close proximity of Mn atoms causes a strong overlap of their \emph{d}-orbitals,
resulting in broad \emph{d}-\emph{d} hybrid bands that cannot support
spontaneous magnetisation. The magnetisation that grows with lattice
expansion can be interpreted as a result of smaller \emph{d}-\emph{d}
overlapping that makes the d-orbitals of Mn atoms more localized.
This \emph{d}-orbital localization then enhances the exchange splitting
of the Mn \emph{d} states reflected in the increased magnetic moments. 

\begin{figure}
\includegraphics[width=8.5cm]{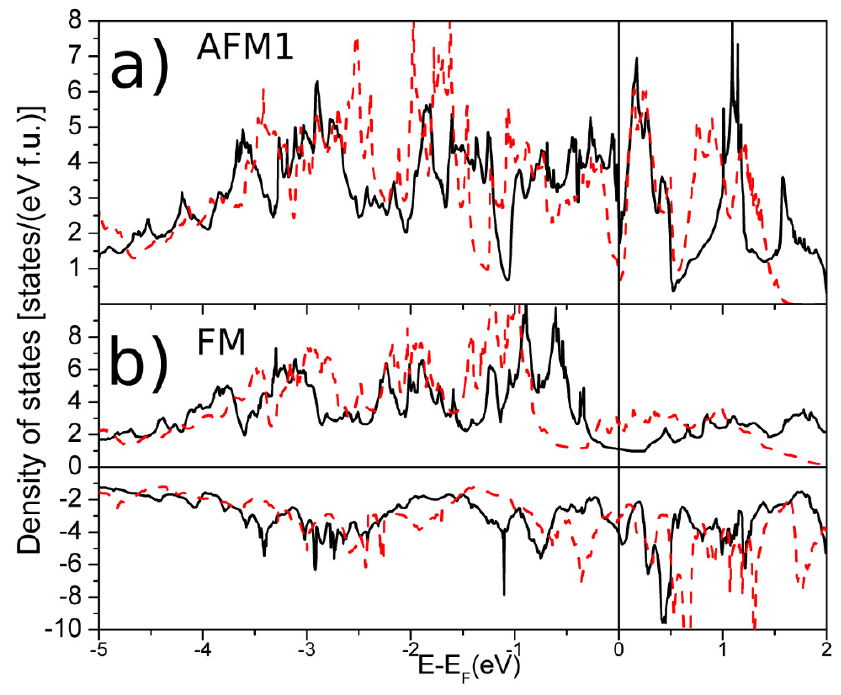}

\caption{\label{fig:MnP-dos} (Color online) Density of states for binary MnP
alloy. The total density of states at the Fermi level ($N_{Tot}(E_{\mathbb{F}})$)
decreases for antiferromagnetic (AFM1) configurations (a) with lattice
expansion from $\varepsilon=0$ (solid line) to $\varepsilon=0.055$
(dashed line), while an opposite tendency is observed in (b) for FM
alignment . }

\end{figure}

In a compressed unit cell the \emph{p}-\emph{d} hybridization of P
and Mn atoms can also lower the \emph{d}-\emph{d} exchange interaction,
reducing the magnetic moments. The enhanced exchange splitting with
increased cell size is demonstrated in Fig.~\ref{fig:MnP-dos}b (bottom),
which shows the total density of states (DOS) of the ferromagnetic
MnP. In the equilibrium volume $(\varepsilon=0)$, the main peaks
of Mn \emph{d}-band in the majority and minority states are separated
by 0.9~eV, which increases to 1.5~eV for the lattice with $\varepsilon=0.055$
lattice expansion. Simultaneously, there is an increase of the total
density of states at the Fermi level, $N_{Tot}(E_{\mathbb{F}})=N_{\Downarrow}(E_{\mathbb{{\normalcolor F}}})+N_{\Uparrow}(E_{\mathbb{{\normalcolor F}}})$
in the FM state from $N_{Tot}(E_{\mathbb{F}})=$ 5.3 to 5.9~states/eV/f.u.
However, a lowering of the DOS of the AFM1 configuration from $N_{Tot}(E_{\mathbb{F}})=5.1$
to 1.8~states/eV/f.u. can be seen, which suggests that the AFM phase
is energetically more favorable Fig.~\ref{fig:MnP-dos}a (top). 

In the next sections, we will show that our stability plot (Fig. \ref{fig:Total-energy_MnP-d1d2}),
calculated for the binary MnP alloy can be used to predict the stability
of AFM or FM states in other (X)MnZ type binary and ternary alloys
with the same (\emph{Pnma}, 62) structure. 


\section{Implications for $XMnP$ AND $Mn(P,Z)$\label{sec:(X)MnP}}

\subsection{$(Fe,Co)MnP$}

A direct extension of the binary MnP alloy is the ternary FeMnP. The
Fe atoms can occupy another 4c site without changing the orthorhombic
crystal symmetry. Interestingly, the addition of iron atoms to the
ferromagnetic MnP results in a commensurate non-collinear AFM structure
with $T{}_{N}=340$~K. The magnetic unit cell is twice as large as
the structural one, being doubled along the c-axis with magnetic moments
of $M{}_{Mn}=2.6\mu_{B}$ and $M{}_{Fe}=0.5\mu_{B}$ lying in the
ab-plane~\cite{FeMnP_Magnetism}. The antiferromagnetism of the system
can be explained by means of the stability plot drawn for MnP in Fig.
\ref{fig:Total-energy_MnP-d1d2}. The added Fe atoms cause an expansion
in lattice dimensions, leading to an increase in both $d_{1}$ and
$d_{2}$ inter-Mn separations into the regime where AFM coupling is
stable. The resulting $d_{1}=3.05$~Å distance is in the proximity
of the region where both AFM1 and AFM3 collinear configurations are
close to the lowest point in total energy, which indicates the trend
to a complex non-collinear AFM ground state. 

The dominant effect of the nearest Mn-Mn distances in determining
the magnetic nature of these alloys becomes more pronounced with the
investigation of the pseudo-ternary (Co,Fe)MnP composition. (Fe$_{\text{1-x}}$Co$_{x}$)MnP
alloys exhibit complex magnetic properties ranging from non-collinear
AFM for $x=0$ through temperature-induced metamagnetism for $0.5\lesssim x\lesssim0.8$
to collinear FM structure above $x\gtrsim0.8$ \cite{CoFeMnP1}. Extensive
theoretical work based on KKR-CPA calculations by Zach et. al. \cite{CoFeMnP1_Dos}
tentatively explained the $N_{Tot}(E_{\mathbb{F}})$ to be partially
accountable for the metamagnetic transition. Their investigation on
various collinear AFM configurations found a formation of a 'pseudogap'
between the conduction and valence electron bands with Co addition,
which results in a lowering of $N_{Tot}(E_{\mathbb{F}})$ similar
to that in the strained MnP alloy of the previous Section (Fig.~\ref{fig:MnP-dos}). 

\begin{figure}
\includegraphics[width=8.5cm]{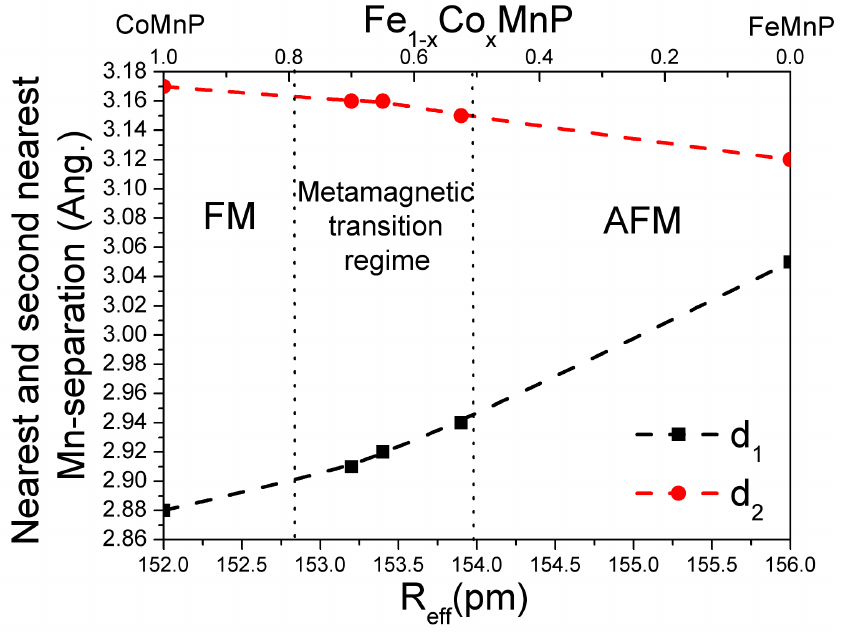}

\caption{\label{fig:CoFeMnP_d1d2} (Color online) Experimentally observed Mn-Mn
distances ($d_{1},d_{2}$) in (Fe$_{\text{1-x}}$Co$_{x}$)MnP pseudo-ternary
alloy at 100~K (after~\cite{CoFeMnP1}). Antiferromagnetism becomes
stable as $d_{1}$ increases above $\sim$2.95Å, in line with our
calculations on MnP in Fig.~\ref{fig:Total-energy_MnP-d1d2}. Temperature-induced
metamagnetism is seen for $0.5\lesssim x\lesssim0.8$.}

\end{figure}

The same authors have also tried to associate the observed magnetic
states to changes in crystal structure. They found the measured Mn-(Fe,Co)
and (Fe,Co)-(Fe,Co) inter-atomic distances to be fairly independent
of temperature. The metamagnetic transition is observed in the region
where 2.90~Å~$\lesssim d_{1}\lesssim$~2.95~Å and 3.15~Å~$\lesssim d_{2}\lesssim$~3.16~Å
with a change much more visible in the $d_{1}$ distance as Fe is
partially replaced by Co when compared to the change in $d_{2}$.
In Fig.~\ref{fig:CoFeMnP_d1d2}, the interatomic Mn-Mn distances
($d_{1},d_{2}$) in the pseudo-ternary (Fe$_{\text{1-x}}$Co$_{x}$)MnP
alloy are plotted as a function of effective atomic radius. Here the
effective atomic radius is defined as $R_{eff}=x\times R_{Co}+(1-x)\times R_{Fe}$,
where $x$ is the concentration, $R_{Co}$ is the atomic radius of
the Co atom (152~pm) and $R_{Fe}$ is the atomic radius of Fe atom
(156~pm), respectively. The metamagnetic transition temperature ($T_{t}$)
shows a strong dependence on Co concentration. It has a value of $T_{t}\thickapprox300$~K
at $x\approx0.5$ and is completely suppressed at $x\approx0.8$~\cite{CoFeMnP1}.
Comparison of the interatomic distances with those seen in the MnP
stability plot of Fig. \ref{fig:Total-energy_MnP-d1d2} shows a remarkably
good agreement between the theoretically expected and the experimentally
found regime of AFM-FM metamagnetism. As the $d_{1}$ interatomic
Mn distance is increased above $\sim$2.95Å the overall structure
is shifted into the region where antiferromagnetism is the dominant
state. This change of magnetism can also be achieved by partial replacement
of the p-block element as demonstrated in the next section. 

\subsection{$MnAs_{1-x}P_{x}$\label{sub:MnAsP}}

The ternary MnAs$_{\text{1-x}}$P$_{x}$ alloy is another example
of metamagnetism and has been extensively studied by Fjellvåg et.
al. in the $0<x<0.20$ composition range~\cite{Fjellvag198429}.
Its rich magnetic and structural phase diagram includes the occurrence
of a magneto-structural transition between hexagonal FM and paramagnetic
(PM) orthorhombic phases around room temperature for $x<0.025$. That
transition, in the related material in Mn$_{\text{1-x}}$Fe$_{x}$As,
recently attracted much attention due to the claims, now disputed,
of an anomalously high MCE ~\cite{MnFePNature2006}. However, the
interest in the origin of the crystal symmetry change is long-standing.
Initially, Goodenough and co-workers associated the transition of
electron states of Mn atoms from high-spin to low-spin to the change
of lattice structure from hexagonal to orthorhombic in Mn(P,As)~\cite{Goodenough1967}.
Recently, Rungger and Sanvito using detailed ab-initio electronic
structure analysis found that if antiferromagnetic alignment in the
basal plane of the hexagonal structure is imposed, the orthorhombic
structure becomes more stable \cite{RunggerMnAs2006}.%
\begin{figure}
\includegraphics[width=8.5cm]{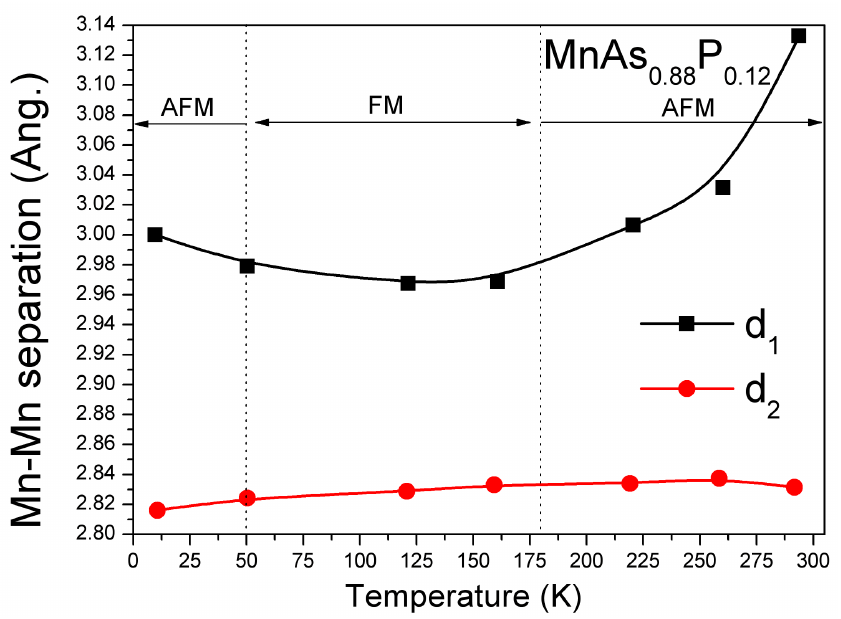}

\caption{\label{fig:MnAsPd1d2} Temperature evolution of the two shortest Mn-Mn
distances ($d_{1},d_{2}$) in ternary MnAs$_{\text{0.88}}$P$_{0.12}$
pseudo-ternary alloy as found experimentally~ \cite{Fjellvag198429}.
In accordance with Fig.~\ref{fig:Total-energy_MnP-d1d2} the AFM
state is stabilised when $d_{1}$ is greater than $~\sim$2.95-2.98~Å.}

\end{figure}

As with MnP case, we focus on magnetism within the orthogonal, \emph{Pnma
}structure. Compositions around $x\thickapprox0.10$ show complex
magnetic behaviour with dominant $H_{a}$-type \cite{Selte1974} helicoidal
AFM spin arrangement at low temperature that reappears at high temperatures
with a transient collinear FM configuration in between. Fjellvag and
co-workers recognised the major importance of the shortest Mn-Mn separations
and used neutron diffraction to map them as a function of temperature.
Fig.~\ref{fig:MnAsPd1d2} reproduces the temperature dependence of
$d_{1}$ and $d_{2}$ that they found in MnAs$_{\text{0.88}}$P$_{0.12}$.
Interestingly, while the shortest Mn-Mn separation stays nearly constant
at around $d_{2}\thicksim$2.85~Å , the $d_{1}$ distance shows remarkable
agreement with Fig.~\ref{fig:Total-energy_MnP-d1d2}: the ferromagnetic
configuration shows stability over the helicoidal AFM one between
50~ K$\lesssim T\lesssim180$~ K where $d_{1}$ stays below 2.95-2.98~Å.
Although a direct extension of our $0$~K stability criteria to finite
temperatures is ambitious, the temperature dependent magnetic behaviour
of MnAs$_{\text{0.88}}$P$_{0.12}$ suggests that the criteria we
have found can be applied more generally.

The materials systems in this section exhibit a broad range of $d_{2}$
values from 2.81~Å in MnAs$_{\text{0.88}}$P$_{0.12}$ to 3.17~Å
in (Fe$_{\text{1-x}}$Co$_{x}$)MnP. Simultaneously, $d_{2}$ changes
from being the shortest to the second shortest Mn-Mn separation. Despite
this change, the variety of magnetic states found in these materials
seem to be well described by critical values of $d_{1}$ only, as
derived from our magnetic stability plot for MnP. In the following
sections, we will further extend our analysis of this plot to XMnZ-type
ternary alloys with the same orthorhombic structure. The importance
of the $d_{1}$ parameter in such materials will also become apparent.

\section{Implications for $XMnZ$-Type Alloys \label{sec:XMnZ}}

\subsection{$XMnSi$ }

\begin{figure}
\includegraphics[width=8.5cm]{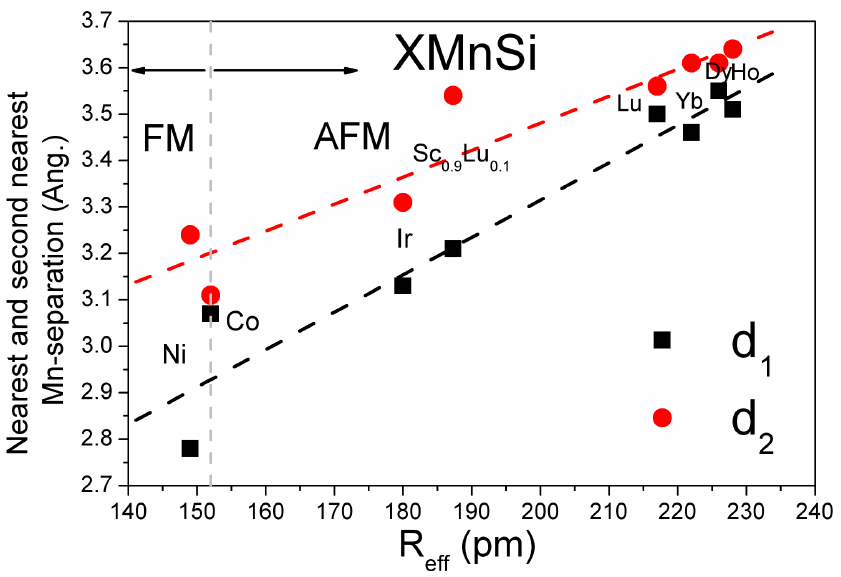}

\caption{\label{fig:XMnSi-Reff} (Color online) Mn-Mn separation ($d{}_{1},d{}_{2}$)
observed in XMnSi ternary alloys, as a function of effective X atomic
radius (low temperature data)\cite{AlexHRPD,Bazela1981,ERIKSSON2005,Ventutini1997,Ijjaali1999}.
Dashed lines are fitted linearly to data and are only a guide to the
eye. }

\end{figure}

There is a fairly large number of XMnSi-type alloys that crystallize
in the \emph{Pnma} orthorhombic structure~\cite{landrum1998,Ventutini1997,Bazela1981,ERIKSSON2005}.
However, many X elements are lanthanoids of large atomic size that
result in considerable expansion of the lattice. The shortest and
second shortest Mn-separations are plotted in Fig.~\ref{fig:XMnSi-Reff}.
The high atomic radius lanthanides cause a large expansion of the
\emph{Pnma} structure and increased Mn-Mn separation, stabilising
the experimentally-found AFM states in accordance with Fig.~\ref{fig:Total-energy_MnP-d1d2}.

The first member of the series in which metamagnetism is observed
is CoMnSi. A field-dependent metamagnetic transition from a AFM helical
state to a high magnetisation state was first reported in this composition
by Nizio{\l} and co-workers\cite{Nizol1978}. The transition temperature
can be dependent on sample preparation, and was found to range from
207~K to 365~K. The peculiar and complex magnetic structure has
been the subject of several investigations \cite{Nizol1978,Niziol1982}.
Early neutron diffraction studies found that a small magnetic moment
is also developed on the Co atoms ($\sim0.3\mu_{B}$) in addition
to the major magnetic contribution from the Mn ($\sim2.2\mu_{B}$)~\cite{Nizol1978}.
The resulting high saturation magnetisation ($\sim$120 emu/g at room
temperature) and field-sensitive metamagnetic transition of the alloy
makes it of interest as a potential room temperature magnetic refrigerant.
We previously reported a magnetic field-induced isothermal entropy
change ($\triangle S$) up to$\sim7$~JK$^{\text{-1}}$kg$^{\text{-1}}$
around room temperature for this alloy \cite{SandemanPRB} and substituted
variants \cite{MorrisonPRB1,MorrisonPRB2}. Recently, unusual negative
thermal expansion along the a-axis of the lattice was also observed
using high resolution neutron diffraction~\cite{AlexHRPD}and in
the absence of an applied magnetic field a crossover in $d_{1}$,
and $\, d_{2}$ is observed. In the above examples, and in those to
follow, the critical value of $d_{1}$ parameter seems to predict
well the change from FM to AFM order (around 2.95~Å). This value,
however, is probably slightly increased in the case of CoMnSi, and
might be associated with the additional exchange competition among
the Co-Co and Co-Mn pairs, compared to the binary MnP-type alloys.
As stated above, Nizio{\l} et. al. identified up to 8 important magnetic
interactions responsible for the overall magnetic properties~\cite{Niziol1982}. 

The end member of the XMnSi series from Fig.\ref{fig:XMnSi-Reff}
is NiMnSi, a ferromagnet with $d{}_{1}=$2.78~Å and $d{}_{2}=$3.24~Å
at 80~K\cite{Bazela1981}. The interpretation of the ferromagnetism
of this ternary is apparent from the previous examples; a structure
that accumulates the smaller Ni atom as a replacement for Co brings
the lattice into the zone where FM coupling of Mn-Mn atoms is more
stable. In the following section we show that metamagnetic behaviour
in more general XMnZ-type \emph{Pnma} structures can also be designed
by the partial replacement of the p-block elements.

\subsection{$NiMnGe_{1-x}Si_{x}$ }

\begin{figure}
\includegraphics[width=85mm]{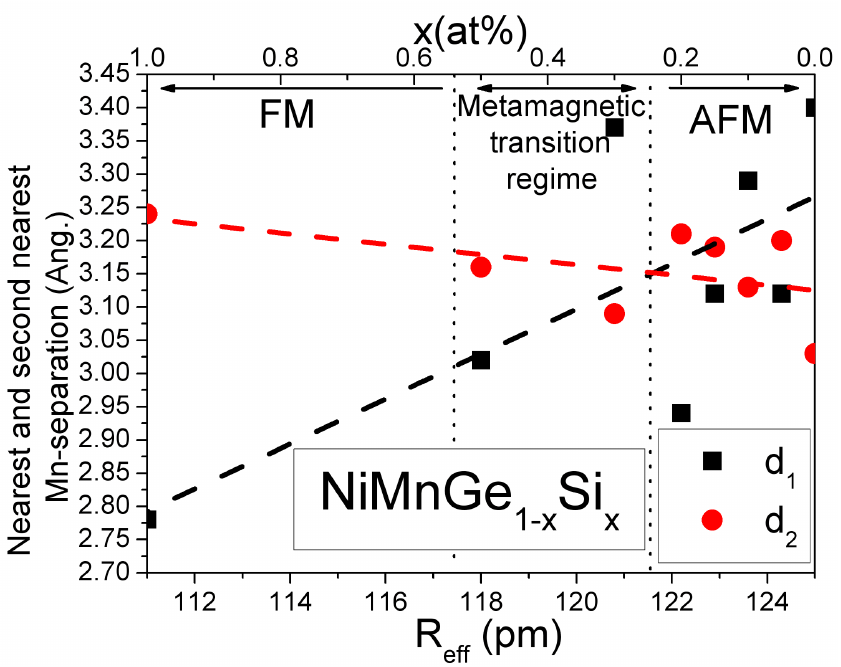}

\caption{\label{fig:NiMnSiGeReef} (Color online) Evolution of the two shortest
Mn-Mn distances ($d_{1},d_{2}$) in pseudo-ternary NiMnGe$_{\text{1-x}}$Si$_{x}$
as a function of composition at 80 K. Field-induced metamagnetism
is reported within the range $0.25\lesssim x\lesssim0.55$ \cite{Bazela1981}. }

\end{figure}

Using the only ferromagnetic example from the previous section, we
now demonstrate that a stable AFM phase can be re-introduced even
in the NiMnSi alloy by the increase of lattice dimensions with larger
p-block elements. Just as with the pseudo-binary MnAs$_{\text{1-x}}$P$_{x}$
in subsection~\ref{sub:MnAsP}, partial replacement of Si by Ge does
not alter the overall valence electron number of the structure. The
NiMnGe$_{\text{1-x}}$Si$_{x}$ system was studied thoroughly by Bazela
et. al.\cite{Bazela1981} and complex magnetic behaviour was observed.
Three regions with different types of magnetic ordering were identified
as a function Si concentration: helicoidal AFM ($0\leq x\lesssim0.25$),
AFM-FM type (NC) metamagnetic ($0.3\lesssim x\leq0.55$) and collinear
FM with $0.55\lesssim x\leq1$. $d_{1}$ and $d_{2}$ are plotted
as a function of Si concentration (top scale) and effective atomic
radus, $R{}_{eff}$ (bottom scale) in Fig.~\ref{fig:NiMnSiGeReef}.
The lack of reported experimental error in the structural parameters
derived from neutron diffraction data\cite{Bazela1981} only allow
us to plot a linear fit line over the data points as a guide to the
eye. The validity of the stability plot calculated for MnP in Fig.~\ref{fig:Total-energy_MnP-d1d2},
however, is once again manifest for such pseudo-ternary NiMnGe$_{\text{1-x}}$Si$_{x}$
compositions: a collinear FM state is stable for $d{}_{1}$ values
below $d{}_{1}\lesssim$2.95~Å and gives way to AFM states for $d{}_{1}\gtrsim$2.95~Å
with increasing effective atomic size (Ge concentration). %

\section{CONCLUSIONS}

We have investigated the occurrence of AFM and FM states in Mn-based
orthorhombic (\emph{Pnma}, 62) alloys. Using DFT theoretical calculations
based on the prototype MnP binary composition, a magnetic stability
plot was constructed taking into account the two shortest Mn-Mn distances
($d_{1},d_{2}$) . We have correlated literature observations of different
magnetic states in Mn-based \emph{Pnma} alloys with this plot and
have found a remarkable agreement. As a result of isotropic expansion
the FM ground state is no longer stable but instead AFM coupling of
the spins on the Mn atoms is predicted when $d{}_{1}\gtrsim$2.95~Å
and $d{}_{2}\gtrsim$2.88~Å, respectively. The $d{}_{1}$ distance
seems to play the dominant role in determining the magnetic state.
In most of the samples the $d{}_{2}$ distance stays fairly constant
as in the case of the (Co,Fe)MnP system or in the temperature dependence
of the MnAs$_{\text{0.88}}$P$_{0.12}$ alloy. In many other cases
$d{}_{2}$ even shows the opposite tendency to $d_{1}$, as seen in
NiMn(Si,Ge). The reason for these phenomena become clearer, if one
examines the \emph{Pnma} crystal structure in~Fig. \ref{fig:MnPstructure}.
The Mn$_{\text{1}}$-Mn$_{\text{2}}$ atoms separated by the distance
$d{}_{2}$ form a continuous chain along the a-axis. This structural
feature prefers spins on these Mn sites to be coupled antiferromagnetically
by superexchange through occupied valence states along Mn-P-Mn. In
fact, the magnetism along the chain probably favours AFM coupling
even below the critical distance of $d_{2}\thickapprox$2.88~Å but
the theoretical calculations are based on an isotropically deformed
lattice and the ratio of the a, b and c parameters for each $\text{\textgreek{e}}$
value have been fixed. Further calculations with anisotropically strained
lattice needs to be carried out in order to clarify this point. The
Mn atoms separated by the nearest-neighbour distance $d_{1}$ (Mn$_{\text{3}}$-Mn$_{\text{4}}$),
on the other hand, do not form a continuous chain in the lattice and
so short-range direct exchange is the dominant interaction. Although
the competition among direct and indirect exchanges that are also
responsible for the observed non-collinearity in many of these alloys
is foreseeable, the overall dominance of the direct exchange of Mn
atoms along the $d_{1}$ separation was not previously evident. The
key role of this separation, resulting in FM coupling for $d_{1}\lesssim$2.95~Å
and AFM coupling above $d{}_{1}\gtrsim$2.95~Å and its control of
the overall magnetism has been shown both theoretically and by comparison
of many examples from the experimental literature. Interestingly,
we find this critical $d{}_{1}$ value, derived from binary MnP, to
be influenced very little by the increased 4c site occupation present
in \emph{Pnma} ternaries. 

Based on these findings, it should be possible to design new alloys
with lattice parameters tuned to a region of competition between AFM
and FM states, enabling field- or temperature-driven metamagnetism.
One example of such a material is Fe$_{\text{x}}$MnP with $x\thickapprox0.6$,
where an Fe deficiency could reduce the lattice volume and bring $d{}_{1}$
into the proximity of FM stability. Based on our findings we believe
that there is still a large number of Mn-based ternary and pseudo-ternary
alloys that can be created with metamagnetic transitions of potential
interest for their magnetocaloric effect. 
\begin{acknowledgments}
The research leading to these results has received funding from the
European Community\textquoteright{}s 7th Framework Programme under
grant agreement 214864. K.G.S. acknowledges financial support from
The Royal Society. Computing resources provided by Darwin HPC and
Camgrid facilities at The University of Cambridge and the HPC Service
at Imperial College London are gratefully acknowledged.
\end{acknowledgments}
\bibliographystyle{apsrev}
\addcontentsline{toc}{section}{\refname}\nocite{*}
\bibliography{MnP_Paper3}

\end{document}